\documentclass[journal]{IEEEtran}
\usepackage{graphicx,times, comment}
\usepackage{amsbsy, amssymb, amsmath, amsfonts, epstopdf}
\usepackage{algorithmic}
\usepackage{algorithm}
\usepackage{color,soul}

\newcommand{\beq}{\begin{equation}}
\newcommand{\eeq}{\end{equation}}
\newcommand{\tbf}{\textbf}
\newcommand{\tit}{\textit}

\newcommand{\ud}{\mathrm{d}}

\newcommand*{\mathcolor}{}
\def\mathcolor#1#{\mathcoloraux{#1}}
\newcommand*{\mathcoloraux}[3]{%
  \protect\leavevmode
  \begingroup
    \color#1{#2}#3%
  \endgroup
}

\newcommand {\Ebb}{\mathbb{E}}
\newcommand{\Ibb}{\mathbb{I}}
\newcommand {\Rbb}{\mathbb{R}}
\newcommand {\Acal}{\mathcal{A}}
\newcommand {\Fcal}{\mathcal{F}}
\newcommand {\Ical}{\mathcal{I}}
\newcommand {\Kcal}{\mathcal{K}}
\newcommand {\Lcal}{\mathcal{L}}

\newcommand {\Pcal}{\mathcal{P}}
\newcommand {\Scal}{\mathcal{S}}

\begin{document}

\title{Multi-Operator Spectrum Sharing for Small Cell Networks : A Matching Game Perspective}

\author{
\IEEEauthorblockN{Tachporn Sanguanpuak\IEEEauthorrefmark{1}, Sudarshan Guruacharya\IEEEauthorrefmark{2}, Nandana Rajatheva\IEEEauthorrefmark{1}, \\ Mehdi Bennis,\IEEEauthorrefmark{1} Matti Latva-Aho\IEEEauthorrefmark{1}}

\IEEEauthorblockA{\IEEEauthorrefmark{1}Dept. of Commun. Eng., Univ. of Oulu, Finland; \IEEEauthorrefmark{2}Dept. Elec. \& Comp. Eng., Univ. of Manitoba, Canada}

\IEEEauthorblockA{Email: \{tsanguan,rrajathe,bennis,matla\}@ee.oulu.fi; guruachs@umanitoba.ca}
}\maketitle

\begin{abstract}

One of the many problems faced by current cellular network technology is the under utilization of the dedicated, licensed spectrum of network operators. An emerging paradigm to solve this issue is to allow multiple operators to share some parts of each others' spectrum. Previous works on spectrum sharing have failed to integrate the theoretical insights provided by recent developments in stochastic geometrical approaches to cellular network analysis with the objectives of network resource allocation problems. In this paper, we study the non-orthogonal spectrum assignment with the goal of maximizing the social welfare of the network, defined as the expected weighted sum rate of the operators. We adopt the many-to-one stable matching game framework to tackle this problem. Moreover, using the stochastic geometrical approach, we show that its solution can be both stable as well as socially optimal. This allows for computation of the game theoretical solution using generic Markov Chain Monte Carlo method. We also investigate the role of power allocation schemes using Q-learning, and we numerically show that the effect of resource allocation scheme is much more significant than the effect of power allocation for the social welfare of the system.

\end{abstract}



\section{Introduction} \label{section:intro}

The next generation 5G cellular network will need to satisfy the performance requirements (e.g., quality-of-service (QoS) and latency) of various applications such as video streaming, data services, and voice communication \cite{Ekram2015}-\cite{Nokia}. In the near future, the total number of existing smart-phones and tablets are projected to be more or less equal to the human population. Most of the devices are expected to be massive machine type communication devices that transmit only a few bytes of data. As such, the spectrum utilization will be an important issue \cite{Nokia}-\cite{Ericsson}. The network operators (OPs) will need to manage their licensed spectrum more efficiently in order to provide service with desired performance requirements \cite{Nokia}.

In recent years, multi-OP spectrum sharing has been gaining attention \cite{Ericsson}-\cite{Kari2014}. Multi-OP spectrum sharing refers to the ability of OPs jointly agreeing on sharing some parts of their licensed spectrum. This approach has emerged as a potential solution to the problem of under-utilization of the dedicated spectrum. The inefficient utilization occurs because the spectrum is often found to be idle at various times. In co-primary (or horizontal) spectrum sharing, OPs have equal ownership of the spectrum \cite{Tim2013}. The spectrum may be shared either orthogonally or non-orthogonally among the OPs. Furthermore, an a-priori agreement should be reached on the spectrum usage with respect to long term sharing of each OP.

\subsection{Related Work} \label{sub-section:related-work}
In \cite{Luca2012}, the authors studied the performance of spectrum sharing in multi-OP LTE networks by using network simulator-$3$ (NS-$3$). They had considered orthogonal spectrum sharing system, where the OPs pool their spectrum with only one OP is allowed to use the spectrum at any given time.
Co-primary spectrum sharing with multiple-input single-output (MISO) and multiple-input multiple-output (MIMO) multiple users in two small cell networks was proposed in \cite{Tachporn2014} and \cite{Tachporn2015}, respectively. The authors considered the case where each base station schedules its users to utilize the shared band when the number of subcarriers in the dedicated band is not enough to serve all users. A matching game based on Gale-Shapley method is proposed for subcarrier allocation. Then, after the users obtain subcarriers, the small cell base stations (SBSs) perform power allocation \cite{Tachporn2014}-\cite{Tachporn2015}. In \cite{Kari2014}, co-primary spectrum sharing in a dense local area was investigated. The authors designed a mechanism to provide flexible spectrum usage between two OPs. The performance of the proposed method was evaluated by system level simulators based on LTE specifications. In \cite{Eduard2014}, the orthogonal spectrum sharing between two OPs was shown to be an important aspect to improve the achievable throughput. The gains in terms of network efficiency is enhanced by sharing spectrum between two OPs. Link level simulation and two hardware demonstrations are given. In \cite{YuTing2012}, a potential game with a learning algorithm was shown to reach the system equilibrium which enhances spectrum efficiency between the OPs. A distributed method is shown to reduce the complexity of inter-OP spectrum sharing, whereas in \cite{Kamal2009}, the same problem is formulated as a two OP non-zero sum game. The utility function of both the OPs was defined by taking the spectrum price and the blocking probability into consideration. Co-primary spectrum sharing was proposed in both centralized and distributed manner in \cite{Petri2015}. System-level simulations for two indoor, small cell layouts were performed. In \cite{PetriICCW2015}, spectrum sharing for multi-operator small cell networks with a guarantee of long term fairness was proposed. Gibbs sampling was studied to develop the decentralized mechanism. A brief version of the current submission is provided in \cite{TachpornICC206}.

\subsection{Contributions and Organization} \label{sub-section:contribution-organization}
\begin{figure}[h]
\centering
\includegraphics[height=3.2 in, width=3.2 in, keepaspectratio = true]{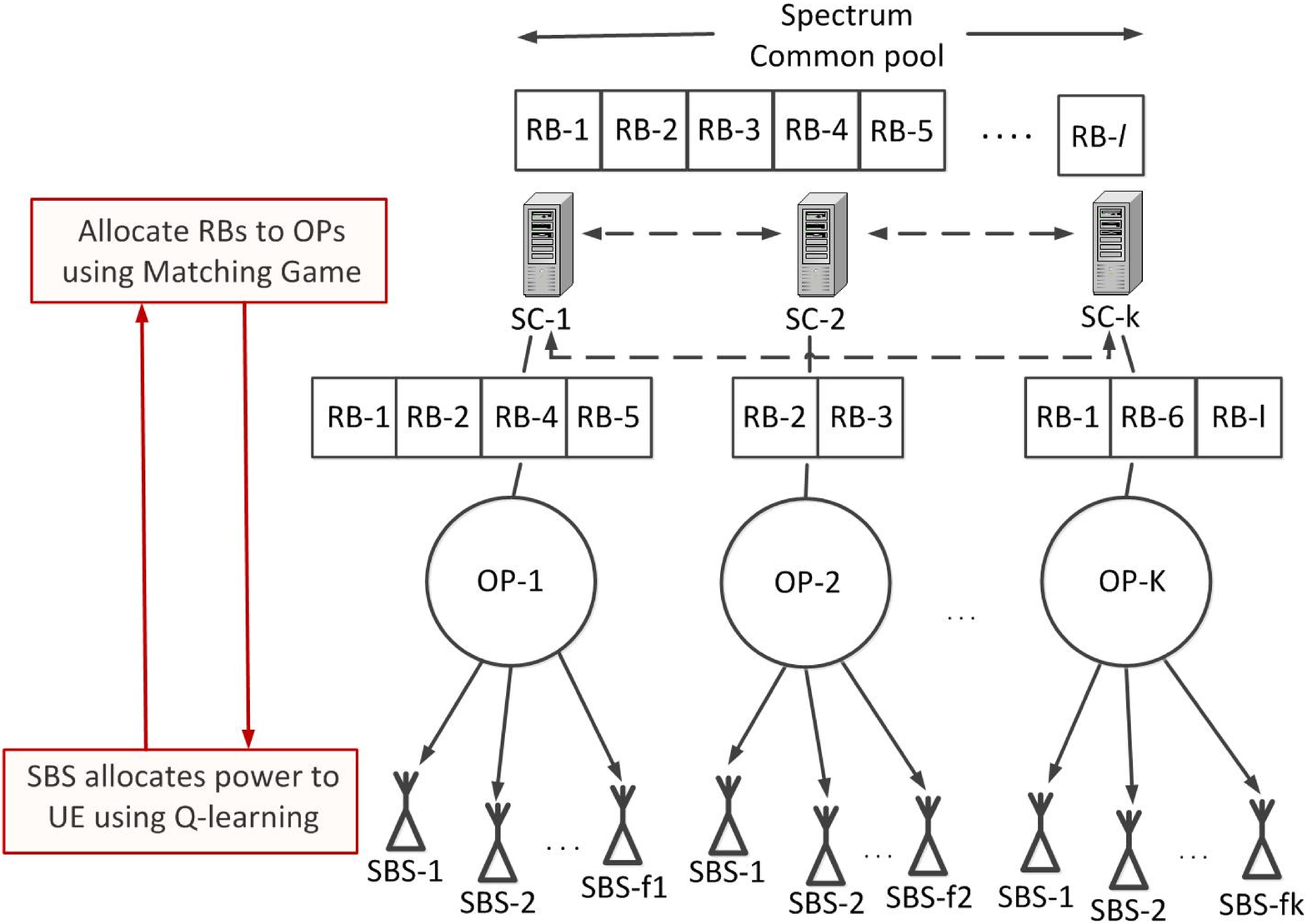}
\caption{Iteration between Matching Game and Q-learning}
\label{fig:matching-Qlearning}
\end{figure}

In this paper, we consider the multi-OP spectrum sharing problem for small cell networks, as shown in Fig. \ref{fig:matching-Qlearning}. Each OP is assumed to serve multiple SBSs, which are spatially distributed according to homogeneous Poisson point process (PPP). Furthermore, each OP connects to a distributed spectrum controller (SC) which is responsible for assigning resource blocks (RBs) from a common pool to the OPs. Here the RB implies one subcarrier. The SC can exchange messages with its components to define a mutual agreement on spectrum sharing policy. We study the problem of spectrum assignment in which the SC can allocate several RBs to an OP, such that multiple OPs can use the same RB.


We summarize the main contributions of this paper as follows:
\begin{itemize}
\item We study the non-orthogonal spectrum assignment with the aim of maximizing the social welfare of the network, defined as the expected weighted sum rate of the OPs. This is essentially a combinatorial optimization problem.  We adopt the many-to-one stable matching game framework to solve this problem. We show that under certain condition, the stable matching solution corresponds to the local optima of the welfare maximization problem.
\item To give a technical proof of this result, we resort to long term average of the network performance in which stochastic geometrical analysis of the expected rate is proposed. Essentially, the average performance depends only on the large scale system parameter like the SBS intensity.
\item The Markov Chain Monte Carlo (MCMC) method is proposed to compute the global solution of social welfare maximization which also leads to the solution of stable matching. The maxima of social welfare can be reached using the MCMC approach, where the sufficient condition given in Corollary 1 is satisfied.
\item In the stochastic analysis as well as the matching game in the operator level, the transmit power of SBS is assumed to be a random variable, but the exact distribution is left undefined. Since the SBS has control over its  transmit power scheme, it is natural to assume that the SBS is interested in maximizing its long term expected data rate by optimizing its power strategy. Thus, we use the Q-learning method to find such an optimal random transmit power scheme for an SBS.
\item From the numerical study, we find that changes in spectrum allocation has a bigger impact on the system performance than changes in power allocation strategy.
\end{itemize}

The rest of the paper is organized as follows. Section \ref{section:systemmodel} describes the system model. Section \ref{section:StoGeoAnaRate} gives a stochastic geometrical analysis of the expected rate of the SBSs. Section \ref{section:BothBSsshareBW} presents inter-OP spectrum sharing using the concept of matching theory. Section \ref{section:Reinforcementlearning} describes Q-learning for random power allocation. The performance evaluation results are presented in Section \ref{section:Numerical Results}. Section \ref{section:Conclusion} states the conclusion.

\section{System Model} \label{section:systemmodel}
\begin{figure}[h]
\centering
\includegraphics[height=6.5 in, width=3.5 in, keepaspectratio = true]{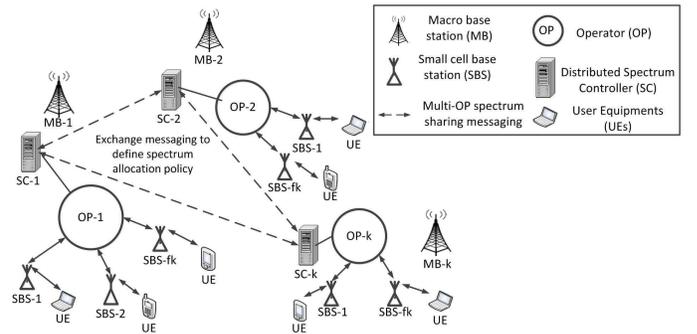}
\caption{System architecture of the multi-OP spectrum sharing}
\label{fig:systemmodel1}
\end{figure}

We propose a multi-OP spectrum sharing for small cell network deployment. The SBSs and the user equipments (UEs) are equipped with single antenna. Fig. \ref{fig:systemmodel1} illustrates the system model under consideration, where SBSs are spatially distributed as homogeneous PPP. The macro base stations (MBs) are assumed to transmit in channels orthogonal to the SBSs; thus, interference from MBs to SBSs is absent. Each OP serves multiple SBSs, and multiple UEs are subscribed to each SBS. Since the system considers single antenna and the SBS is assumed to employ time division multiple access (TDMA) scheme, hence each SBS can serve only one UE in a given time slot. The problem of user scheduling is not the main concern and is beyond the scope of this paper. Nevertheless, the analysis does not lose its generality, since we tend to consider an ``average'' UE, in the sense that we ultimately average over the random channel gains as well as the random distance of the UE from the SBS. 

The spectrum of OPs serving the SBSs is assumed to be divided into dedicated bands and a shared band. The dedicated band of an OP can be allocated only to the SBSs associated with the given OP, while the shared band can be accessed by multiple OPs and can be allocated to their respective SBSs. Since the dedicated spectrum of each OP is assumed to be fixed and predetermined, our study focuses only on allocating the shared spectrum to the OPs.

Consider a set of multiple OPs given by $\Kcal$ with $K$ OPs. Let the set of SBSs subscribed to an OP-$k$ be given by the $\Fcal_k$ with $F_{k}$ SBSs. We assume that each OP has the same spatial intensity of SBSs per u nit area. Also, let $\Fcal = \cup_{k\in\Kcal} \Fcal_k$ be the set of all SBSs. Since each SBS is assumed to serve a single UE at a given time slot, for an SBS-$f$ in $f\in\Fcal$, we will denote its associated UE by UE-$f$ at any given time. The channel state information is assumed to know at each SBS. The set of RBs in the shared band available to the network is given by $\Lcal$ with $L$ RBs. Let $\Lcal_k \subset \Lcal$ be the set of RBs assigned to OP-$k$ with $L_k$ RBs.

The SBSs associated with OP-$k$ can select any one of the RBs in $\Lcal_k$ to serve its UE. We assume that the SBS will uniform randomly choose a single RB from $\Lcal_k$. Hence, an SBS's transmit power is restricted to a single RB. Let the total power of each SBS be given by $p_{tot}$, which is discretized into $N = \frac{p_{tot}}{\delta}$ levels, where $\delta$ is a quanta of power. Thus, the set of transmit power levels that an SBS-$f$ can choose from is $\Pcal_f = \{0, \delta, 2 \delta, \ldots, (N-1) \delta \}$. We shall denote the transmit power of the SBS-$f$ by $p_f \in \Pcal_f$. The SBSs are assumed to use a probabilistic scheme to select a suitable power level $n \in \{0, 1,\ldots,N-1\}$. Thus, any given action taken by an SBS can be simply represented by $n$.

We consider the analysis of a single time slot of the overall system, so the utility of SBSs has been calculated as the expected data rate when a single UE is present. We further assume that any RBs allocated to an OP can be accessed by more than one SBS associated with that OP. Thus, the expected rate of the UE-$f$ associated with SBS-$f$ is given by
\begin{equation}
R_f = \Ebb \Big[ \log_2\Big( 1 + \frac{h_{ff}^{(l)} r_{ff}^{-\alpha} p_{f}}{\sum_{f'\in\mathcal{I}_l} h_{f'f}^{(l)} r_{f'f}^{-\alpha} p_{f'} + \sigma^2 }\Big)\Big],
\label{eqn:rate-AP}
\end{equation}
where $p_{f}$ is the transmit power of SBS-$f$ on RB-$l$, $h_{f'f}^{(l)}$ is the channel fading gain between UE-$f$ and SBS-$f'$ using RB-$l$. For simplicity, we assume the fading to be Rayleigh. Also, $\alpha$ denotes pathloss exponent and $r_{f'f}$ is the distance between the UE-$f$ and SBS-$f'$. The $\mathcal{I}_l \subset \Fcal$ is the set of SBSs using the same RB-$l$, while $\sigma^2$ is the noise variance. Here, the expectation is taken with respect to the random channel gains, the random distance of the UE from the SBS, as well as probabilistic channel access and power allocation strategy. We can interpret this averaging effect as averaging over multiple time slots as SBS serves different UEs in different locations with differing channel gains.

The interference experienced by a UE of an SBS can be categorized as either intra-OP interference or inter-OP interference. The intra-OP interference is caused by the fact that the SBSs associated with a given OP can access any RB assigned to that OP. Thus two SBSs served by the same OP can access the same RB. On the other hand, the inter-OP interference is caused by the fact that a given RB can be shared by two or more OPs.

The expected data rate of OP-$k$ will be the sum of expected rates of each SBS. We can express the rate of OP-$k$ as,
\begin{equation}
R_{OP_k}(\Fcal_k, \Lcal_k) = \sum_{f\in\Fcal_k} \rho_{f} R_f,
\label{eqn:rate-OP}
\end{equation}
where $\rho_{f}$ denotes the weight of each SBS which is a positive real number.

\section{Preliminary Analysis of Expected Rate} \label{section:StoGeoAnaRate}
In this section, we deal with the analysis of the expected rate of an SBS, where we explicitly consider the randomness due to channel fading, distance geometry, and random channel access. Thus, the only source of randomness we will not analyze is the one due to the power allocation method. In this section, we assume that a form of random power allocation method is available, then it will be updated with the optimal random power allocation scheme using Q-learning method to be explained in Section \ref{section:Reinforcementlearning}.

Let the rate of a generic downlink SBS-UE system transmitting in a fixed RB-$l$ and at fixed power level $n$ be given by,
\begin{equation}
R_n^{(l)} = \log(1 + SINR_n^{(l)}).
\label{eqn:rate-AP-l-n}
\end{equation}

Here the $SINR_n^{(l)}$ is given by,
\begin{equation}
SINR_n^{(l)} = \frac{h_{ff}^{(l)} r_{ff}^{-\alpha} p_{f}}{\sum_{f'\in\mathcal{I}_l} h_{f'f}^{(l)} r_{f'f}^{-\alpha} p_{f'} + \sigma^2 }.
\label{eqn:SINR-AP-l-n}
\end{equation}

Taking the expectation of $R_n^{(l)}$ with respect to the channel gains and interference, we get $\Ebb[R_n^{(l)}] = \Ebb_{h_{ff}^{(l)},I_f^{(l)}} \Big[ \log\Big( 1 + \frac{h_{ff}^{(l)} r_{ff}^{-\alpha} p_{f}}{I_f^{(l)} + \sigma^2 } \Big) \Big],$
where $I_f^{(l)}=\sum_{f'\in\mathcal{I}_l} h_{f'f}^{(l)} r_{f'f}^{-\alpha} p_{f'}$ is the interference experienced by UE-$f$ in RB-$l$. Using the fact that, the expectation of any positive random variable $x$ can be given by $\Ebb[x] = \int_0^\infty P(x>t) \ud t$, the integral becomes $\Ebb[R_n^{(l)}] = \Ebb_{I_f^{(l)}} \Big[ \int_{t=0}^\infty P\Big( h_{ff}^{(l)} > \frac{r_{ff}^\alpha  (e^t - 1)(\sigma^2 + I_f^{(l)})t}{p_f} \Big) \ud t \Big].$
Since we have assumed Rayleigh fading, $h_{ff}^{(l)}$ is exponentially distributed (i.e. $h_{ff}^{(l)} \sim \exp(\eta)$). Thus, we have $\Ebb[R_n^{(l)}] = \Ebb_{I_f^{(l)}} \Big[ \int_{t=0}^\infty \exp \Big\{ \eta \frac{r_{ff}^\alpha  (e^t - 1)(\sigma^2 + I_f^{(l)})t}{p_f} \Big\} \ud t \Big].$
Putting $\nu = \frac{\eta r_{ff}^\alpha  (e^t - 1)t}{p_f}$, we can simplify the expression as, $\Ebb[R_n^{(l)}] = \int_{t=0}^\infty \exp(-\nu \sigma^2) \cdot \Ebb_{I_f^{(l)}} [\exp (-\nu I_f^{(l)})] \ud t.$
Here, we can recognize the expectation with respect to the interference as the Laplace transform of $I_f^{(l)}$ as $\mathfrak{L}_{I_f^{(l)}}(\nu) = \Ebb_{I_f^{(l)}} [\exp (-\nu I_f^{(l)})].$
Thus, we have
\begin{equation}
\Ebb[R_n^{(l)}] = \int_{t=0}^\infty \exp(-\nu \sigma^2) \mathfrak{L}_{I_f^{(l)}}(\nu) \ud t.
\label{eq:R-n-l}
\end{equation}

For the homogeneous PPP $\Phi$, the moment generating functional of a two dimensional ($2$D) space PPP $\Phi$ can be obtained by taking the Laplace transform of $\sum_{x\in\Phi} f(x)$ of intensity $\lambda$ using the formula \cite{Baccelli2009}, $\mathfrak{L} (\nu) = \Ebb [\exp( - \nu \sum_{x\in\Phi} f(x))] = \exp\{-\lambda \int_{\Rbb^2} (1- e^{-\nu f(x)}) \ud x \}.$

Let $\lambda_{k,l}$ denote the spatial intensity of SBSs served by OP-$k$ and using RB-$l$. By the merging property of a Poisson process, the total intensity of SBSs under OP-$k$ is $\lambda_k = \sum_{l\in\Lcal_k} \lambda_{k,l}$. Also, when the RBs are uniform randomly selected, via thinning property of Poisson process, we have $\lambda_{k,l} = \lambda_k/L_k$. Again, via the merging property of Poisson process, the total intensity of interfering SBSs utilizing an RB-$l$ is given by
\begin{equation}
\lambda_l = \sum_{k \in \Ical_l^{OP}} \lambda_{k,l} = \sum_{k \in \Ical_l^{OP}} \frac{\lambda_k}{L_k},
\label{eqn:intensity-inteference}
\end{equation}
where $\Ical_l^{OP} \subset \Kcal$ is the set of OPs allocated with RB-$l$
Since we assumed that each OP has the same spatial intensity of SBSs per unit area, we have $\lambda_k = \lambda$.

For our case, the $\lambda_l$ from (\ref{eqn:intensity-inteference}) is the required intensity of the interfering SBSs per unit area. For a given RB-$l$, this is the sum of individual intensities of all OPs transmitting in that RB. Thus, we have
\begin{align*}
 \mathfrak{L}_{I_f^{(l)}}(\nu) =&  \Ebb_{I_f^{(l)}} [\exp (-\nu I_f^{(l)})] = \exp\{- \lambda_l \int_{\Rbb^2} (1 - \\
                                              &  \Ebb_{h_{f'f}^{(l)}} \Ebb_{p_{f'}} [\exp(-\nu p_{f} r_{f'f}^{-\alpha} h_{f'f}^{(l)})]) \ud A_{f'f}\}.
\end{align*}


For $2$D space, $\ud A_{f'f} = 2\pi r_{f'f} \ud r_{f'f}$, so
\begin{align*}
\mathfrak{L}_{I_f^{(l)}}(\nu) = & \exp\{- 2 \pi \lambda_l r_{f'f} \int_0^\infty (1 - \\
                                           &  \Ebb_{h_{f'f}^{(l)}}\Ebb_{p_{f'}} [\exp(-\nu p_{f} r_{f'f}^{-\alpha} h_{f'f}^{(l)})]) \ud r_{f'f}\}.
\end{align*}

Following \cite{Semasinghe2015}, this expression can be simplified to obtain
\begin{equation}
\mathfrak{L}_{I_f^{(l)}}(\nu) = \exp\{-\pi \lambda_l \Ebb_{p_{f'}} [p_{f'}^{2/\alpha}] \; \Ebb_{h_{f'f}^{(l)}} [ h_{f'f}^{(l) 2/\alpha} ] \nu^{2/\alpha} \Gamma(1-2/\alpha)\},
\label{eq:mathfrak}
\end{equation}
where $p_{f'}$ is the transmit power of an SBS in the set of interferers and $\Gamma(z)$ is the complete Gamma function. By substituting the expression (\ref{eq:mathfrak}) in (\ref{eq:R-n-l}), we get
\[ \Ebb[R_n^{(l)}] = \int_0^\infty \exp ( -\nu \sigma^2) \exp (- C \nu^{2/\alpha} ) \ud t, \]
where $C = \pi \lambda_l  \Ebb_{p_{f'}} [p_{f'}^{2/\alpha}] \; \Ebb_{h_{f'f}^{(l)}} [ h_{f'f}^{(l) 2/\alpha} ] \Gamma(1-2/\alpha)$.

Since dense small cell networks will be interference limited, we can neglect the noise term $\exp(-\nu \sigma^2)$ in the integral for the expected rate, so the only significant term to be integrated is $\Ebb[R_n^{(l)}] = \int_0^\infty \exp(- C \nu^{2/\alpha}) \ud t.$
Furthermore, when we take $\alpha = 4$, we obtain an analytically tractable form of $\Ebb_{h_{f'f}^{(l)}} [ \sqrt{h_{f'f}^{(l)}} ]$ as given in \cite{Semasinghe2015},
$\Ebb_{h_{f'f}^{(l)}} [ \sqrt{h_{f'f}^{(l)}} ] = \frac{1}{2}\sqrt{\frac{\pi}{\eta}}.$
Thus, assuming $\eta=1$ and substituting the expression for $\nu$, we get the equation for the expected rate as
\beq
\Ebb[R_n^{(l)}] = \int_0^\infty \exp \Big( \frac{-\lambda_l \pi^2 r_{ff}^2 \Ebb[\sqrt{p_{f'}}]}{2\sqrt{p_f}} \sqrt{(e^t - 1)t}  \Big ) \ud t.
\label{eq:rateSBS}
\eeq

Note that this formula for $\Ebb[R_n^{(l)}]$ is independent of the number of interfering SBSs in RB-$l$. The expected rate depends only on the intensity of SBSs $\lambda_l$, the probability mass function (PMF) of the interferers selecting transmit power of level $n$ in RB-$l$, and the actual transmit power of SBS-$f$. The PMF of $p_{f'}$ can either be interpreted as the percentage of SBSs transmitting in RB-$l$ with power level $n$ or the distribution resulting from some random power selection method.

If we assume that the UE associated with the SBS is located uniform randomly around a circular area of radius $r_c$ with the SBS as the center, then $f_R(r_{ff}) = \frac{2r_{ff}}{r_c^2}$, for $r_{ff} > 0$. De-conditioning on $r_{ff}$, we have
\[ \Ebb[R_n^{(l)}] = \frac{1}{\pi \lambda_l r_c^2} \int_0^\infty \ud t \int_0^\infty 2 \pi \lambda_l r_{ff} \exp (- B \pi \lambda r_{ff}^2) \ud r_{ff}, \]
where $B = \frac{\pi \Ebb[\sqrt{p_{f'}}]}{2\sqrt{p_f}} \sqrt{(e^t - 1)t}$. We can integrate integral to the right as $\int_0^\infty 2 \pi \lambda_l r_{ff} \exp (- B \pi \lambda r_{ff}^2) \ud r_{ff} = 1/B$. Substituting the expression for $B$, we have
\begin{equation}
\Ebb[R_n^{(l)}] = \frac{\Ebb[\sqrt{p_{f'}}]}{2 \lambda_l r_c^2 \sqrt{p_f}} \int_0^\infty \sqrt{(e^t - 1)t} \ud t.
\label{eq:rateSBS-2}
\end{equation}


\tit{Discussions:}
\begin{enumerate}
\item The above analysis holds for any generic SBS located at any location. This is guaranteed by the Slivnyak's theorem \cite{Baccelli2009}, according to which the statistics for the PPP is independent of the test location. This also implies that the SBSs transmitting over an RB-$l$ are identical. That is, every SBS will experience same interference statistics.


\item Since the SBS-$f$ can access any one of $L_k$ RBs assigned to its associated OP-$k$ with equal probability of $1/L_k$, we can express $R_f$ in (\ref{eqn:rate-AP}) in terms of $R_n^{(l)}$ in (\ref{eqn:rate-AP-l-n}) as,
\begin{equation}
R_f = \Ebb_{n,l}[\Ebb[R_n^{(l)}]] = \frac{1}{L_k} \sum_{l\in\Lcal_k} \Ebb_n [\Ebb[R_n^{(l)}]].
\label{eqn:rate-AP-2}
\end{equation}
For fixed PMFs of $p_f$ and $p_{f'}$, the expression $\Ebb_n [\Ebb[R_n^{(l)}]]$ will be some constant dependent on the value of $\lambda_l$ for the RB-$l$. Thus, the average rate $R_f$ depends only on the intensity of interfering SBSs $\lambda_l$ in RB-$l$ and the number of available RBs $L_k$.

\item For the important special case, when $L_k=1$ for all $k\in\Kcal$, we will have from (\ref{eqn:intensity-inteference}), $\lambda_l = \sum_{k \in \Ical_l^{OP}} \lambda_k$. When we further assume constant intensities of SBS for all OP, $\lambda_k = \lambda$, then $\lambda_l = |\Ical_l^{OP}| \lambda$. In this case, it is clear that $R_f$ will depend only on $|\Ical_l^{OP}|$, the number of OPs assigned to the same RB-$l$, provided that the PMF of both $p_f$ and $p_{f'}$ remain fixed.
\end{enumerate}

\section{Multi-Operator Spectrum Sharing using Matching Game} \label{section:BothBSsshareBW}
\subsection{Problem Statement} \label{subsection:ProblemStatement}
Consider the social welfare of the network to be the overall weighted sum rate as follows:
\begin{equation}
S(\mu) = \sum_{l\in\mathcal{L}} \sum_{k\in\Kcal} x_{lk} \rho_k R_{OP_k}(\Fcal_k, \Lcal_k),
\label{eq:socialwelfare1}
\end{equation}
where $\tbf{X} = |\mathcal{L}|\times|\Kcal|$ is a matching matrix $\{x_{lk} : (l,k) \in \mathcal{L} \times \Kcal \}$. We denote the matrix $\tbf{X}$ as,
\begin{equation}
x_{lk} = \left\{
\begin{array}{cl}
1 & \mbox{iff}\; \mu(\mbox{OP}_k) = \mbox{RB}_l \\
0 & \mbox{otherwise},
\end{array}
\right.
\end{equation}
where $\mu$ is a matching. Lastly, $\rho_k$ is the weight of OP-$k$.

The objective of the matching game for the multi-OP spectrum sharing is to maximize the social welfare. Thus, the optimization problem can be expressed as,
\begin{align}
& S^*(\mu) = \max_{\tbf{X}} \sum_{l\in\mathcal{L}} \sum_{k\in\Kcal} x_{lk} w_k R_{OP_k}(\Fcal_k, \Lcal_k), \nonumber\\
\mbox{s.t.} \quad\quad &  \mbox{(C1)} \quad \sum_{l\in\Lcal} x_{lk} \leq b_l  \quad \forall l \in \mathcal{L}, \label{eqn:centralproblem} \\
& \mbox{(C2)} \quad \sum_{k\in\Kcal} x_{lk} \leq c_k \quad \forall k \in \Kcal.  \nonumber
\end{align}

Condition $\mbox{(C1)}$ assures that each RB-$l$ can be allocated to at most $b_{l}$ OPs, and condition $\mbox{(C2)}$ guarantees that each OP-$k$ gets at most $c_k$ RB. We will refer to $c_k$ as the  resource ``demand'' of OP-$k$, while $b_l$ as the resource ``supply'' of $l\in\Lcal$. This is a binary integer programming problem. In this paper, we try to solve it by providing a method based on a game theoretic model known as matching games. We will use the framework of many-to-one matching game with externality which is described as follows.

\subsection{Matching Theory for Multi-Operator Spectrum Sharing}\label{section:Matching game}

Since the OPs are assumed to be able to communicate with each other through a distributed SC during the spectrum sharing process, it makes sense to use the cooperative game theory to model the decision process of the OPs. Here we will describe our solution approach using the matching game theory, which is an instance of cooperative game theory. Unlike the traditional matching games, which are based on preference relationships defined between two sets of players, we will be following a more recent framework described in \cite{Baron2011}, which directly deals with utilities rather than preferences. In our case, the sets of players can be written as, ($\mathcal{K}, \mathcal{L}$). In our application, we have slightly modify the approach given in \cite{Baron2011} to suit our purpose.

The kind of matching game we are interested in is complicated by the presence of externalities. Normally, it is assumed that the preference of a player does not depend on the other players' preferences. This assumption does not fit in our multi-OP spectrum sharing problem. The choice made by an OP-$k$ to select an RB will affect the data rate of other OP-$k'$, where $k'\neq k$, that transmits over the same RB. This in turn will determine the desirability of the RB for the other OP-$k'$. Hence, the preferences of each OP over the set of RB depends on the particular matching $\mu$. In game theory, such a phenomenon is known as an externality. These are external effects that dynamically change the performance of each OP. Thus, the proper framework to deal with our problem is to utilize matching games with externalities \cite{Baron2011},\cite{Roth1992}.

Theoretically, our problem fits into the framework of many-to-many matching with externality. However, since many-to-many matching with externality is still a topic of ongoing research, we will adapt the better understood framework of many-to-one matching with externality for our work. In many-to-one framework, a matching $\mu$ describes the assignment of OPs to RBs such that one OP can be matched to only one RB, whereas one RB can be matched to multiple OPs.

\begin{figure}[h]
\centering
\includegraphics[height=2.5 in, width=3 in, keepaspectratio = true]{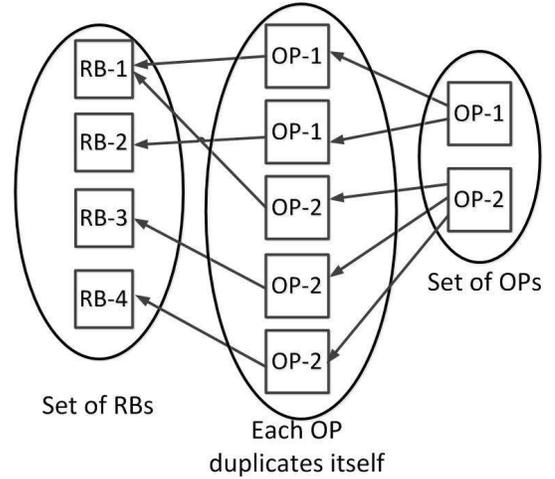}
\caption{Matching between RB and OPs}
\label{fig:matchingOP-RB}
\end{figure}

One  further complication with this approach is that the direct application of existing many-to-one matching algorithms would result in allocation of at most a single RB to an OP. However, our problem allows us to allocate more than one RB to an OP, as given by constraint (C2) in ($\ref{eqn:centralproblem}$). To tackle this problem, we create an augmented set of players by producing identical copies of OPs, as shown in Figure \ref{fig:matchingOP-RB}. Each copy of OP inherits all the SBSs associated with its OP-$k$, $k\in\Kcal$, which we shall refer to as the parent OP. Let $\Kcal_k = \{k_1, \ldots, k_{c_k}\}$ denote the set of identical copies of OP-$k$, which we shall refer to as children OPs of parent OP-$k$. Thus, our augmented set of OPs is $\Kcal_{aug} = \cup_{k\in\Kcal} \Kcal_k$. Since each child OP is assigned with at most one RB in many-to-one matching, if the number of children OPs is equal to the the resource demand of the parent OP-$k$, $c_k$, then this method guarantees that each parent OP can obtain more than one RB. At the same time, by allocating at most one RB to each child OP, it ensures that each parent OP will get at most the maximum number of allowed RBs.

However, it requires that the children of OP-$k$ in the group of players $\Kcal_k$ coordinate with each other such that no two players in $\Kcal_k$ selects the same RB. Otherwise, each parent OP will be assigned with a lower number of RBs than the requirement. We illustrate these ideas in Fig.  \ref{fig:matchingOP-RB}. Here, OP-$1$ requires two RBs, so it makes two copies of itself; whereas OP-$2$ requires three RBs, so it makes three copies of itself.

For a given parent OP-$k$, we will take the rate of SBS and child OP to be given by (\ref{eqn:rate-AP-2}) and (\ref{eqn:rate-OP}), respectively. Note that for SBSs associated with child OP, the number of RB, $L_{k'}$, where $k'\in\Kcal_k$, used in (\ref{eqn:rate-AP-2}) is at most unity. Thus the special case given in the third point of Discussion in Section \ref{section:StoGeoAnaRate} applies. Since the children OP inherits all the SBS of its parent OP, we will need to take the average rate of parent OP with respect to the number of available RBs, as
\begin{equation}
R_{OP_k} = \frac{\sum_{k' \in \Kcal_k} R_{OP_{k'}}}{L_k}.
\end{equation}

To address these concerns, we will use the idea of swap matching as described in \cite{Baron2011}, which considers peer effects of a social network and a weaker notion of stability, known as two-sided exchange stability. This model is distinguished by the use of utility functions rather than traditional preference ordering. The authors of \cite{Baron2011} have applied their idea to student-hostel matching problem. In our work, the students are represented by OPs while the hostels are represented by RBs. Thus, we propose a decentralized approach that can guarantee the maximum number of RBs required for each OP, while at the same time ensuring that each RB is not utilized by more than the limited number of OPs.

\subsection{Many-to-One Matching with Externalities} \label{section:Groupstablematching}
In this part, we will describe the framework of many-to-one matching with externalities, with some modifications for our purpose. More formally, we can describe a matching as:

\textbf{Definition $1$} :  For many-to-one matching, a matching is a subset $\mu \subseteq \mathcal{L} \times \Kcal_{aug}$ such that $|\mu(k)| = 1$ and $|\mu(l)| = b_{k}$ where $\mu(k)= \{l \in \mathcal{L} : (l,k) \in \mu\}$ and $\mu(l)= \{k \in \mathcal{K}_{aug} : (l,k) \in \mu\}$.

Also, for any $k\in\Kcal_{aug}$, let $\mu^2(k)$ denote the children of the same parent OP-$k$ who utilize the same RB-$l$. We will denote the desirability of RB-$l$ for any OP-$k$ by $D_l^k \in \Rbb^+ \cup \{0\}$. In our case, the desirability of an RB for children OPs is given by the weighted sum rate obtained by the OP, when it accesses that RB as given in (\ref{eqn:rate-OP}). For a given matching $\mu$, we can write the desirability as $D_{\mu(k)}^k$.
The utility of OP-$k$ is given by,
\beq
U_k(\mu) = D_{\mu(k)}^k \cdot \Ibb_{\mu}(k),
\eeq
where the indicator function $\Ibb(\cdot)$ is given by
$ \Ibb_{\mu}(k) = \left\{\begin{array}{rl} 0 & \mbox{if} \; \mu^2(k) \neq \emptyset \\
							 1 & \mbox{otherwise.}
			 \end{array} \right. $

In other words, if two children of the same parent OP access the same RB, they will be punished. This has the effect of ensuring that two sibling OPs will access different RBs. Here we have modified the definition of utility given in {\cite{Baron2011}} by using the product of desirability with an indicator function, instead of defining the utility as the sum of desirability and a penalty term.


A \tit{swap matching} $\mu_{k}^{k'}$ is a matching $\mu$ in which the OPs $k$ and $k'$ switch places while keeping all assignments of other OPs the same. More formally:

\tbf{Definition $2$} : Given a matching $\mu$, a swap matching $\mu_{k}^{k'} = \{ \mu \backslash\{(k,l),(k',l')\} \} \cup \{ (k,l'), (k',l) \}$.

The players involved in the swap are two OPs and two RBs. The two OPs switch their respective RBs while all other assignments remain the same. In this framework, it is possible that one of the OPs involved is a ``hole,'' representing an available vacancy in RB that an OP can move to fill in. When two actual OPs are involved, this type of swap is a two-sided version of ``exchange.''

Two-sided exchange stability requires that the two OPs involved approve the swap. Here, we give a slightly modified version of its definition.

\textbf{Definition $3$} : A matching $\mu$ is two-sided exchange-stable (pairwise stable) iff there does not exist a pair of OPs $(k, k')$ such that
\begin{enumerate}
\item $\forall i \in \{k,k'\}, U_{i}(\mu_k^{k'}) \geq U_{i}(\mu)$,
\item $\exists i \in \{k,k'\}, U_{i}(\mu_k^{k'}) > U_{i}(\mu)$, and
\item $\forall i \in \{\mu(l), \mu(l') \}\backslash \{k,k'\}, U_{i}(\mu_k^{k'}) \geq U_{i}(\mu)$.
\end{enumerate}

In other words, a swap matching in which all OPs involved are indifferent is called two-sided exchange-stable. Also, a swap is approved if both OPs involved in a switch experience an improvement in their utilities, with at least one OP doing strictly better than before. A ``hole'' will always be indifferent. We have modified the definition given in {\cite{Baron2011}} by adding a third condition which states that, for the approval of the swap, all the OPs occupying the RBs involved in the swap should see an improvement in their utilities as well.

\subsection{Stability of Many-to-one Matching with Externalities}\label{section:Existencematching}
In this part, we will show the existence of the many-to-one stable matching with externalities for multi-OP spectrum sharing. We will prove that all local maximas of the social welfare are pairwise stable. We first define what we mean by local maxima, and then give a few lemmas, after which we will prove our theorems.

First, let the potential of the system be defined as,
\beq
\phi(\mu) = \sum_{k\in\Kcal_{aug}} D_{\mu(k)}^k \Ibb_{\mu}(k).
\eeq

\textbf{Definition $4$}: The local maximum of the potential $\phi(\mu)$ is the matching $\mu$ for which there exists no matching $\mu'$ which is obtained from $\mu$ by swapping any two OPs $k, k'$ such that $\phi(\mu') > \phi(\mu)$.

We now show that the desirability of RB-$l$ for the rest of the OPs that use this RB-$l$, and which are not involved in a swap process, either improves or remains unchanged after the swap has occurred.

\tbf{Lemma $1$} : For any swap matching $\mu_k^{k'}$, $D_{\mu_k^{k'}(j)}^j \geq D_{\mu(j)}^j$ for all $j \in \Kcal_{aug}\backslash \{k, k'\}$.

\tbf{Proof} : Since each OP in $\Kcal_{aug}$ utilizes only a single RB, we can invoke the third point in the Discussion given in Section \ref{section:StoGeoAnaRate}. There are three possible cases.
First, for all OP-$j$ not assigned to RB-$l$ or RB-$l'$ (i.e., $\mu(j) \notin \{l, l'\}$), the number of OPs associated with its RB, $\mu(j)$, does not change; hence its  $D_{\mu_k^{k'}(j)}^j = D_{\mu(j)}^j$.
Second, assuming (without loss of generality) that OP-$k'$ is a ``hole'', for all OP-$j$ assigned to RB-$l$, after the swap, the number of OPs on RB-$l$ decreases. So, its $D_{\mu_k^{k'}(j)}^j > D_{\mu(j)}^j$.
Lastly, For all OP-$j$ assigned to either RB-$l$ or RB-$l'$ (i.e., $\mu(j) \in \{l, l'\}$), after the swap, the number of OPs on RB-$l$ and RB-$l'$ remain the same. Thus, $D_{\mu_k^{k'}(j)}^j = D_{\mu(j)}^j$. $\square$

\textbf{Lemma $2$} : Any swap matching $\mu_k^{k'}$ such that,
\begin{enumerate}
\item $\forall i \in \{k,k'\}, U_{i}(\mu_k^{k'}) \geq U_{i}(\mu)$,
\item $\exists i \in \{k,k'\}, U_{i}(\mu_k^{k'}) > U_{i}(\mu)$, and
\item $\forall i \in \{\mu(l), \mu(l') \}\backslash \{k,k'\}, U_{i}(\mu_k^{k'}) \geq U_{i}(\mu)$.
\end{enumerate}
leads to $\phi(\mu_k^{k'}) > \phi(\mu)$.

\tbf{Proof} : The difference in potential between two matching is given by  $\phi(\mu_k^{k'}) - \phi(\mu) =  \sum_{i \in \Kcal_{aug}} [ D_{\mu_k^{k'}(i)}^i \Ibb_{\mu_k^{k'}}(i) - D_{\mu(i)}^i \Ibb_{\mu}(i) ].$ According to Lemma $1$, $D_{\mu_k^{k'}(i)}^i = D_{\mu(i)}^i$ for all $i \in \Kcal_{aug} \backslash  \{\mu(l), \mu(l')\}$. That is, the desirability of an RB-$l$ for the rest of the OPs occupying RBs other than RB-$l$ and RB-$l'$ does not change after the swap. This also means that for these OPs, $\Ibb_{\mu_k^{k'}}(i) = \Ibb_{\mu}(i)$. Then the difference in potential is only due to the differences in the desirability of the OPs occupying the RBs involved in the swap: $\phi(\mu_k^{k'}) - \phi(\mu) =  D_{l'}^k \Ibb_{\mu_k^{k'}}(k) - D_l^k \Ibb_{\mu}(k) + D_l^{k'} \Ibb_{\mu_k^{k'}}(k') - D_{l'}^{k'} \Ibb_{\mu}(k') + \sum_{j \in \{\mu(l), \mu(l') \}\backslash \{k,k'\}}  [ D_{\mu_k^{k'}(j)}^j \Ibb_{\mu_k^{k'}}(j) - D_{\mu(j)}^j \Ibb_{\mu}(j) ].$


Assuming that the conditions (1), (2) and (3) of the lemma are satisfied, then without loss of generality, assume that the performance of OP-$k$ strictly improves. Then the change in utility of OP-$k$ is $0 < U_k(\mu_k^{k'}) - U_k(\mu)  = D_{l'}^k \Ibb_{\mu_k^{k'}}(k) - D_l^k \Ibb_{\mu}(k).$
Similarly, for OP-$k'$, we have $0 \leq U_{k'}(\mu_k^{k'}) - U_k(\mu) =  D_l^{k'} \Ibb_{\mu_k^{k'}}(k') - D_{l'}^{k'} \Ibb_{\mu}(k').$
 Lastly, for OP-$j$, where $j \in \{\mu(l), \mu(l') \}\backslash \{k,k'\}$, we have $0 \leq U_{j}(\mu_k^{k'}) - U_j(\mu) = D_{\mu_k^{k'}(j)}^j \Ibb_{\mu_k^{k'}}(j) - D_{\mu(j)}^j \Ibb_{\mu}(j).$
Adding the inequalities, we have $0 < \phi(\mu_k^{k'}) - \phi(\mu),$ proving our lemma. $\square$

In the following theorems, the Theorem $1$ ensures the existence of an optimal matching, while Theorem $2$ ensure that this matching is pairwise-stable.

\textbf{Theorem $1$}: There exists at least one optimal matching.

\tbf{Proof} : This is easily seen to be the case since the number of matching is finite. Thus, there must exist at least one optimal matching  which leads to the maximum social welfare. $\square$

\textbf{Theorem $2$} : All local maxima of $\phi$ are pairwise stable.

\tbf{Proof} : Let the matching $\mu$ be the local maximum of $\phi(\mu)$. Lemma $2$ shows that any swap matching that satisfies both conditions in Lemma $2$ strictly increases the overall social welfare. If there exists another swap matching $\mu'$, then this contradicts the assumption that $\mu$ is a local maximum. Thus, $\mu$ must be stable. $\square$

\textbf{Corollary $1$} : If $\Ibb_{\mu}(k) = 1$ for all $k\in \Kcal_{aug}$, then all local maxima of the system objective $S$ are pairwise stable.

\subsection{Swap Matching Algorithm}\label{section:matching-algo}

Computationally, the swap matching can be performed by any pair of OPs by calculating their own local utilities and swapping their obtained RBs with each other based on mutually beneficial conditions given in Lemma $2$. Thus, in theory, the swap matching can be implemented distributively by comparing the local utilities of the pairs of OPs, without the need for a central controller. The problem with this approach is that the system can be stuck in a local optima.

However, as stated in Theorem $2$ and its corollary, all local maxima of the social welfare are also pairwise stable, under certain condition. In other words, the solution to the swap matching problem corresponds to the solution of the social welfare maximization problem. It also means that we can alternatively compute the stable matching solutions by computing the solution to a global social welfare function. Since finding the maxima of the social welfare is a  combinatorial problem, the maxima of social welfare can be reached using the generic approaches. 

Algorithm \ref{alg:mcmc} proceeds to optimize the social welfare $S$ via the Markov Chain Monte Carlo (MCMC) method. We first initialize with a random matching, and at each iteration, we proceed to accept a swap of random pair of OPs based on the probability that depends on the change in social welfare. It keeps track of the best matching found thus far. We can give a greedy version of the Algorithm \ref{alg:mcmc} by removing the exploration steps given in the lines $9$ to $11$. The resulting Greedy Swap Algorithm proceeds in a greedy fashion to improve the social welfare, and it is possible to implement it distributively. Since the social welfare strictly improves with each iteration, this algorithm converges to a two-sided exchange-stable matching.

In the MCMC and Greedy Swap approaches, the sufficient condition of Corollary $1$ is  satisfied by carefully selecting the pair of RBs and OPs to be swapped. The MCMC approach is efficient in the sense that it enables us to find a better optima by giving the system a chance to overcome a local optima in which it can be stuck. The computation of such global objective function necessitates a centralized system. Nevertheless, both approaches lead to the same solution of social welfare maximization.

\begin{algorithm}
\caption{MCMC Swap Algorithm}
\label{alg:mcmc}
 \begin{algorithmic}[1]
  \STATE Initialize the matching matrix $\mathbf{X}$.
  \STATE Compute the initial data rate of each OP-$k$.
 	\FORALL{$t \leq $ maxIterations}
 		\STATE Select a random pair of RBs $\{l, l'\}$.
        \STATE Search for OPs $\{k,k' \in K_{aug}\}$ using the RBs $\{l, l'\}$, respectively.
        \STATE Swap the two RBs for each OP $\{k, k' \}$ to obtain $\mu_{k}^{k'}$.
        \STATE Update the expected rate of augmented OPs \{$k, k'$\}, $k,k' \in K_{aug}$, with the Q-learning.
        \STATE Compute the social welfare $S_{t}(\mu)$ in (\ref{eq:socialwelfare1}).
        \STATE Compute the transition probability $P_{T_b} = \frac{1}{1 + e^{-T_{b}(S(\mu_k^{k'}) - S(\mu))}}$.
            \IF {$rand() < P_{T_b}$}
            \STATE  $\mu \leftarrow \mu_k^{k'}$ and $S_{t}(\mu) = S_{t}(\mu_k^{k'})$
            \ELSIF{$S_{t}(\mu) > S_{t-1}(\mu_{k}^{k'})$}
 		    \STATE $\mu \leftarrow \mu_k^{k'}$
            \STATE Update the social welfare $S_{t}(\mu) = S_{t}(\mu_k^{k'})$
 	        \ENDIF
        \STATE $t \leftarrow t+1$.
    \ENDFOR
 \end{algorithmic}
 \end{algorithm}

\section{Power Allocation for Smallcell Base Stations using Q-Learning Strategy}\label{section:Reinforcementlearning}
In our paper, we have discretized the transmit power levels of SBS and have assumed that the SBS transmits by accessing any one of the power levels by some fixed randomization scheme. Recall that welfare maximization problem (\ref{eq:socialwelfare1}) formulated among the OPs in Section \ref{section:BothBSsshareBW} does not regard the transmit power of SBSs as one of the optimization parameters. As such, any randomization scheme would have been sufficient for the purpose of the welfare maximization and the swap matching process among the OPs. Similarly, in Section \ref{section:StoGeoAnaRate}, we dealt with the analysis of the expected rate of an SBS, where we implicitly assumed that some random power allocation method was available, although it was left undefined, from which we could calculate the expected rate, with respect to the random power. 

In this section, we investigate the optimal transmit power scheme. Since an SBS has control over its  transmit power, it is natural to assume that the SBS is interested in maximizing its long term expected data rate by optimizing its power strategy. However, the optimal probability mass function (PMF) defined over the discrete power levels is not known a-priori by the SBS. Thus, we will use Q-learning to find such an optimal power PMF for an SBS. 

The Q-learning method is a distributed algorithm which relies only on local information available at each SBS. Hence, there is no information exchange and coordination among SBSs. We assume that all the SBSs are able to estimate the interference they experience on each RB and accordingly tune their transmission strategies towards a better performance. With this ability to learn, each SBS-$f$, $f\in\Fcal_k$, belonging to OP-$k$,  where  $k\in\Kcal_{aug}$, uses the RB allocated to OP-$k$ to serve its corresponding UE based on Q-learning.


The $Q$-learning model consists of a set of states $\Scal$ and actions $\Acal$ aiming at finding a policy that maximizes the observed rewards over the interaction time of the agents/players. For our case, the agents are the SBSs. Every SBS $f \in \Fcal_k$ served by an OP-$k$, where $k \in \Kcal_{aug}$  explores its environment, observes its current state $s$, and takes a subsequent action $a$, according to a decision policy $\psi: s \rightarrow a$. With their ability to learn, the knowledge about other players' strategies is not needed. Instead, a $Q$-function preserves what they have learned from their interaction with other players in the network, based on which, better decisions can be made.

For each OP-$k$ belonging to the set $k\in\Kcal_{aug}$, let us denote by $\mathcal{G}_k^{Q}=\big(\Fcal_k,\lbrace \mathcal{P}_f \rbrace_{f\in \Fcal_k},\lbrace w_f \rbrace_{f\in \Fcal_k}\big)$ the $Q$-learning game. Here, the players of the game are the SBSs $f \in \Fcal_k$ which seek to allocate power in the RBs assigned to its corresponding OP. The $s_f(t)$ is the state of SBS-$f$ at time $t$. The state of an SBS is a binary variable, $s_f(t) \in \{0,1\}$, which indicates whether SBS-$f$ experiences interference in RB-$l$ assigned to its corresponding OP-$k$ such that its required QoS is violated. The QoS requirement is said to be violated when $SINR_n^{(l)} < SINR_{th}$, where $SINR_n^{(l)}$ is given by (\ref{eqn:SINR-AP-l-n}). The $a_f(t)$ is the action of SBS-$f$, where $a_f(t) \in \Pcal_f$. Any given action can be represented by an integer variable $a_f (t) \equiv n$, where $n$ represents the power level. Finally, $w_f(t)$ is the reward or payoff of SBS-$f$ at time-instant $t$, which we take as the instantaneous rate of SBS-$f$ at time-instant $t$ as given by (\ref{eqn:rate-AP-l-n}) if the QoS is satisfied, otherwise it is taken to be zero:
\begin{equation}
w_f(t) = \left\{
\begin{array}{cl}
R_n^{(l)} & \mbox{iff}\; SINR_n^{(l)} \geq SINR_{th} \\
0 & \mbox{otherwise}.
\end{array}
\right.
\label{eq:SINRnl}
\end{equation}

The \emph{expected} discounted reward over an infinite horizon is given by:
\begin{equation}
 V^{\psi}(s)=\Ebb  \Big[ \gamma^t \times w(s_t,\psi^*(s_t))|s_0=s \Big],
\label{eqn:V}
 \end{equation}
where $0\leq \gamma \leq 1$ is a discount factor and $w$ is the agent's reward at time $t$. Equation (\ref{eqn:V}) can be rewritten as:
\begin{equation}
 V^{\psi}(s)=W(s,\psi^*(s)) +\gamma \sum_{v \in S} P_{s,v}(\psi(s))V^{\psi}(v),
\label{eq:VI}
 \end{equation}
where $W(s,\psi^*(s))=\Ebb\{w(s,\psi(s))\}$ is the mean value of reward $w(s,\psi(s))$, and $P_{s,v}$ is the transition probability from state $s$ to $v$. Moreover, the optimal policy $\psi^*$ satisfies the optimality criterion:
\begin{equation}
 V^{*}(s)= V^{\psi^*}(s)=\max_{a\in \mathcal{A}}\left(W(s,a) + \gamma \sum_{v \in \mathcal{S}} P_{s,v}(a)V^{*}(v)\right),
\label{eq:VII}
 \end{equation}

It is generally difficult to explicitly calculate the reward $W(s,a)$ and transition probability $P_{s,v}(a)$. However, through $Q$-learning, the knowledge of these values can be gradually learnt and reinforced with time. For a given policy $\psi$, we can define a $Q$-value as:
 \begin{equation}
 Q^*(s,a)=W(s,a)+\gamma \sum_{v \in S} P_{s,v}(a)V^{\psi}(v),
 \label{eq:VIII}
 \end{equation}
which is the discounted reward when executing action $a$ at state $s$ and then following policy $\psi$ thereafter.

Here, we use the $Q$-learning algorithm to iteratively approximate the state-action value function $Q(s,a)$. The agent keeps trying all actions in all states with non-zero probability and must sometimes explore by choosing at each step a random action with probability $\epsilon \in (0,1)$, and the greedy action with probability $(1-\epsilon)$. This is referred to as $\epsilon$-greedy exploration \cite{Dusit2009}, \cite{Mehdi2011}. Another option is to use the Boltzmann exploration strategy with temperature parameter $T_p$ \cite{Fudenberg1998}, where the action $a$ in state $s$ is taken with a probability $P(a|s)$, and the SBS receives a reinforcement $w$. The actions are chosen according to their $Q$-values as:
\begin{equation}
 P(a|s)= \frac{e^{Q(s^k,a)/T_p}}{\sum_{a' \neq a}e^{Q(s^k,a')/T_p}}.
 \end{equation}

The $Q$-learning process aims at finding $Q(s,a)$ in a recursive manner where the update equation is given as \cite{Mehdi2011}:
\begin{align}
Q_{t+1}(s_t,a_t) = & (1-\beta_t)Q_{t}(s_t,a_t) + \nonumber \\
                        & \beta_t \left[w(s_t,a_t)+ \gamma \max_{a_{t}'\neq a_t}Q_{t}(s_t,a_{t}')\right],
\label{eqn:Qupdate}
\end{align}
where $\beta_t$  is the learning rate, such that $0\leq\beta_t<1$. The Q-learning algorithm for power allocation at each SBS-$f$ is described in the Algorithm \ref{alg:Qlearning}.

\begin{algorithm}
\caption{$Q$-learning algorithm for power allocation}
\label{alg:Qlearning}
\begin{algorithmic}[1]
 \STATE $Q(s,a)=0$
        \FORALL {$t \leq $ maxIterations}
        \FOR {$k = 1:K_{aug}$}
        \STATE Calculate the utility $u_f$
            \IF {$rand() \leq \gamma$}
            \STATE Randomly choose an action (power level) $n$
            \ELSE
            \STATE Choose a state with $n^{*} = \text{argmax}_{n}Q(s,a)$
            \ENDIF
        \STATE Each SBS-$f$ computes the expected date rate ($R_f$).
        \STATE Update $Q$-value $Q_{t+1}(s_t,a_t)=(1-\beta_t)Q_{t}(s_t,a_t)+ \beta_t \left[w(s_t,a_t)+ \gamma \max_{a_{t}'\neq a_t}Q_{t}(s_t,a_{t}')\right]$.
        \STATE $t \leftarrow t+1$
        \ENDFOR
        \ENDFOR
 \end{algorithmic}
 \end{algorithm}

\subsection{Convergence of Q-Learning} \label{section:Convergence_Qlearning}
The optimal Q-function is a fixed point of a contraction operator $\mathbf{H}$, defined for a function $Q : \Scal \times \Acal \to \Rbb$ for the decision policy $\psi$ as,
\begin{equation}
\mathbf{H}_{Q}(s,a) = W(s,a) + \gamma \sum_{v \in S} P_{s,v}(a)\max_{a' \neq a} Q(s,a').
\label{eqn:Hq-definition}
\end{equation}
This operator is a contraction in the sup-norm \cite{Fudenberg1998} such that,
\begin{equation}
||\bold{H}_{Q_{1}} - \bold{H}_{Q_{2}}||_{\infty} \leq \gamma ||Q_{1} - Q_{2}||_{\infty}
\label{eqn:Hq-contraction}
\end{equation}

With any initial estimate $Q_{0}$, the Q-learning uses the update rule as in (\ref{eqn:Qupdate}) to converge to an optimal decision policy. To show the convergence proof of Q-learning algorithm, we need the following auxiliary result from stochastic approximation \cite{Fudenberg1998,Jaakkola1994}:

\textbf{Theorem $3$} :  Let a random process $\{\Delta_t\}$, taking values in $\Rbb^{n}$, be defined as $\Delta_{t+1}(s) = (1-\beta_t)\Delta_t(s) + \beta_t F_t(s),$
then $\Delta_t$ converges to zero with probability one under the assumptions:
\begin{enumerate}
\item $0 \leq \beta_t \leq 1, \sum_{t=1}^{\infty}\beta_t = \infty$ and $\sum_{t=1}^{\infty}\beta^{2}_{t}< \infty$
\item $||\Ebb[F_t(s)|| \leq \gamma||\Delta_t||$, with $\gamma \in (0,1)$
\item var[$F_t(s)] \leq C(1+ ||\Delta_t||^{2})$ for $C$ is some constant.
\end{enumerate}

\textbf{Theorem $4$} : The Q-function converges to its optimal value with probability one under the condition that: $0 \leq \beta_t \leq 1$, $\sum_{t=1}^{\infty}\beta_t = \infty$, $\sum_{t=1}^{\infty}\beta^{2}_{t} < \infty$.

\tbf{Proof} : Following the update equation of Q-learning in (\ref{eqn:Qupdate}), subtracting $Q^{*}(s_t, a_t)$ from both sides and letting $\Delta_t( s_t,a_t)= Q_{t}(s_t,a_t) - Q^{*}(s_t,a_t)$ yields, $\Delta_{t}(s_t, a_t) = (1-\beta_t)\Delta_t(s_t,a_t) + \beta_t[w(s_t, a_t) + \gamma \max_{a' \neq a} Q_{t}(v,a') - Q^{*}(s_t,a_t)]$.

Let $F_{t}(s,a)$ be given by $F_{t}(s,a) = w(s, a, X(s, a)) + \gamma \max_{a' \neq a}Q_t(v,a') - Q^{*}(s,a),$
where $X(s,a)$ is a random sample state obtained from the Markov chain $(\Scal, P_{s,v}(a))$. Taking the expectation of $F_t$, we have $\Ebb[F_{t}(s,a)] = W(s,a) + \sum_{s\in S}P_{s,v} [\gamma \max_{a'\neq  a}Q_t(v,a') - Q^{*}(s,a)] = \mathbf{H}_{Q_{t}}(s,a) - Q^{*}(s,a).$
Here, the second equality follows from the definition of $\mathbf{H}_Q$ as given in (\ref{eqn:Hq-definition}), while $\sum_{s\in S}P_{s,v} Q^{*}(s,a) = Q^{*}(s,a)$, since $Q^{*}(s,a)$ is a constant. The fixed point due to contraction operator $\mathbf{H}$ leads to $Q^{*} = \bold{H}Q^{*}$. Thus, we can re-express $\Ebb[F_{t}(s,a)] = \mathbf{H}_{Q_{t}}(s,a) - \mathbf{H}_{Q^*}(s,a).$
Now from the contraction property of $\mathbf{H}_Q$ given in (\ref{eqn:Hq-contraction}), $||\Ebb[F_{t}(s,a)]||_\infty  \leq \gamma||Q_t - Q^{*}||_{\infty} = \gamma||\Delta_t||_\infty$.
This verifies the second assumption given in Theorem $3$.

Since the reward in our case is the rate of UE-$f$ associated with SBS-$f$, as given by (\ref{eqn:rate-AP}), the reward is a bounded, deterministic function. This ensures that the third assumption given in Theorem $3$ is also confirmed, as shown in \cite{Jaakkola1994}.

Thus, by the Theorem $3$, $\Delta_{t}$ converges to zero with probability one. That is, $Q_{t}$ converges to $Q^{*}$ with probability one. $\square$
%
%

\section{Numerical Results}\label{section:Numerical Results}
In this section, we present numerical results to evaluate the performance of our multi-OP spectrum sharing framework and proposed algorithms. The system is iteratively updated as shown in Fig. \ref{fig:matching-Qlearning}. The SBSs are spatially distributed according to homogeneous PPP inside a $500$ meters radius of circular area. Moreover, each OP is assumed to have the same intensity of SBSs per unit area. We assume the intensity of SBS to be $8/(\pi \times 500^2)$ per square meter. Each SBS serves a single UE and each UE is located within $20$ meters of the SBS. For $K$ OPs, let the resource demand made by each OP be given by the vector $\tbf{c} = [c_1,\ldots,c_K]$. The vector $\tbf{c}$ also tells us how many children of each parent OP will there be in the augmented OP set. For simplicity, we assume the weights in the social utility function to be $\rho_f= \rho_k = 1$.
The direct pathloss between SBS and SBS-UE at distance $d$ meters is given by $PL(d) = 37+20 \log_{10}(d)$ dB, and the pathloss due to the wall ($PL_{\text{wall}}$) is $15$ dB. The standard deviation of log-normal shadow fading is assumed to be $4$ dB. The cross-gain pathloss between SBS and SBS-UE at distance $d_{\text{S-UE}}$ is given by $PL(d_{\text{S-UE}}) = 7 + 56 \log_{10}(d_{\text{S-UE}}) + PL_{\text{wall}}$. The maximum transmit power of each SBS is $10$ dBm, and the noise variance is $-120$ dBm. The SINR threshold at each user is $3$ dB. The temperature $T_b$ in MCMC algorithm is set to be $100$. In the Q-learning algorithm, we set the parameters as: discount factor $\gamma = 0.95$, exploration probability $\epsilon = 0.1$, learning rate $\beta_t = 0.5$. In all the cases, the summation for all elements of $\tbf{c}$ is kept less than or equal to number of RBs assigned to the OP-$k$ i.e. $\sum_{k \in \Kcal}c_k \leq \sum_{l \in \Lcal} b_l$, in order to ensure that the total resource demand is less than the total supply of resources. Also, we keep $c_k \leq L$, to ensure that the resource demand of each OP is always fulfilled.

Unless otherwise stated, for the stochastic averaging during the simulations, we have taken the ensemble average from 2000 instances of random geometric configurations of 8 SBS per OP, scattered uniform randomly over a circular area of radius 500 meters. Also the swap algorithms were run for 2000 iterations.

\subsection{Convergence of the Swap Algorithms}

\begin{figure}[h]
\centering
\includegraphics[height=5.9 in, width=3.8 in, keepaspectratio = true]{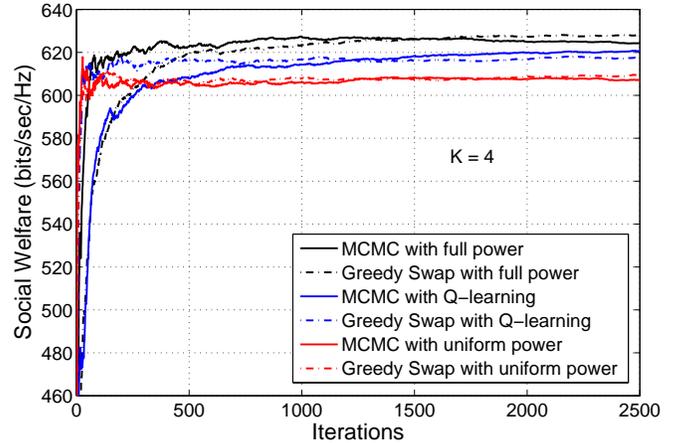}
\caption{The convergence of social welfare for $K= 4$ OPs using MCMC and greedy swap algorithms with full power allocation, power allocation using Q-learning, and uniform power allocation}
\label{ConvergenceMCMCgreedy}
\end{figure}

In Fig. \ref{ConvergenceMCMCgreedy}, the convergence of the social welfare using MCMC and greedy swap algorithms is given when there are $K=4$ OPs with full power allocation, power allocation using Q-learning, and uniform power. We fix $\tbf{c}=[4,4,4,4]$ for $K=4$, $L=6$, and $b_l=4$ and the running average of the social welfare is plotted against the iteration. In this figure, we run 2500 iterations for the swaps algorithms. We see that the system converges to a  steady state. We also notice that for a given power allocation scheme, both the MCMC and greedy swap algorithms converge to similar steady state performance. This demonstrates that both the swap algorithms are equally effective.

At the steady state, the full power allocation achieves the highest average social welfare, while the uniform power allocation achieves the lowest average social welfare. The difference in  steady state performance between full power and uniform power allocation schemes is about 20 bps/Hz. The reason why the Q-learning power allocation scheme does not perform better than full power allocation scheme is that there is always a non-zero probability for the Q-learning to visit a less than optimal power allocation method during its exploration step.

We also note that for full power allocation, the greedy swap algorithm takes longer time to converge to the steady state than MCMC swap algorithm. Similarly, for power allocation using Q-learning, the greedy swap algorithm converges faster than the MCMC swap algorithm. Lastly, for uniform power allocation, both the greedy swap as well as MCMC swap algorithm converge at similar rate. However, since the operation of the swap algorithm does not depend on the underlying power allocation algorithm, it would be incorrect to associate these differences in speed  due to the power allocation algorithm being used. This means that, given our data, it remains inconclusive as to which of the two swap algorithms is faster.

\subsection{Effect of Changing the Number of Operators and Power Allocation Scheme}
%
%
\begin{figure}[h]
\centering
\includegraphics[height=6.1 in, width=3.8 in, keepaspectratio = true]{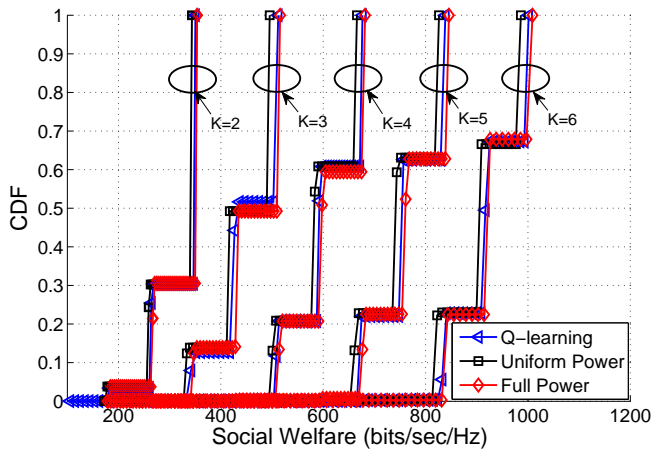}
\caption{Comparison of the cumulative distribution function (CDF) of social welfare for $K=2,3,4,5$ and $6$ OPs using MCMC algorithm with full power allocation, power allocation using Q-learning, and uniform power allocation}
\label{CDFOP2to6}
\end{figure}

In Fig. \ref{CDFOP2to6}, we plot the cumulative distribution function (CDF) of the overall social welfare (bits/sec/Hz) using MCMC algorithm for different numbers of OPs and different power allocation schemes. We fix the number of available RBs to $L=6$ and the number of OPs that can utilize the same RB-$l$ is $b_l=4$ for all $l \in \Lcal$, and $c_k = 4$ for all $k\in\Kcal$. We consider cases when each SBS allocates power to its UE using uniform power allocation, Q-learning, and full power allocation. A few observations are as follows:

1) The CDF of social welfare occurs as discontinuous steps. This can be explained by the fact that during the search for optimal matching using Greedy Swap Algorithm or MCMC Swap Algorithm, the system gets stuck in a number of locally optimal solutions. The percentage of time spent in each local optima is given by the height of the step. For example, when $K = 2$, the system spend around 70\% of its time in a solution that gives around 350 bps/Hz, while the system spends less than 30\%
 of its time in a solution that gives around 250 bps/Hz.

2) The height of the last step, which represents the best locally optimal solution obtained within a fixed number of iterations, is seen to be decreasing as the number of OPs increases. This means that the system tends to spend less time in that state as the number of OPs increase. This can be explained by the fact that as the number of OPs increase, the number of possible matching of the system increases, but since we have used only a fixed number of iterations to run the swap algorithm, this shortens the amount of time spent in the best solution.

3) For a given number of OPs, we see that the difference in  performance caused by differing power allocation scheme  is much smaller than the change in performance caused by differing matching scheme. In other words, for a given number of OPs, when we look at the family of CDF curves for various power allocation schemes, the difference in the width of the step of the CDF curve, caused by the system's transition to a new matching scheme, is much larger than the difference  in performance due to differing power allocation scheme for a particular matching. For example, when $K=3$ the difference in performance, when the locally optimal matching is changed, is roughly 100 bps/Hz, as given by the width of the steps of its CDF curve. This is in contrast to the difference in performance due to power allocation scheme when the system is in a given matching, which for $K=3$, is about 25 bps/Hz. Thus, the effect of resource allocation scheme is much more significant than the effect of power allocation scheme for the social welfare of the system.

4) The difference in the median values of the CDFs tends to remain more or less constant as the number of OPs is increased. This is because for $L=6$, when there are three or more OPs in the system, the system becomes interference limited. As such, the average social welfare obtained per OP is similar, as can be seen in Fig. \ref{AvgwelfareperOP} for $L=6$. Thus as the number of OPs increase, the median tends to increase linearly. This is in contrast to the case when $K=2$. For this case, since $\tbf{c} = [4,4]$ and $L=6$, the best configuration is where each OP has two RBs that it does not share with the other OP. Thus, out of four required RBs, two RBs are free from inter-operator interference while the other two RBs are not free from inter-operator interference. This leads to a higher average social welfare for each OP. Thus, the median of the CDF for $K=2$ is closer to the median of the CDF of $K=3$.

\subsection{Effect of Changing the Number of Resource Blocks}

\begin{figure}[h]
\centering
\includegraphics[height=5.9in, width=3.8 in, keepaspectratio = true]{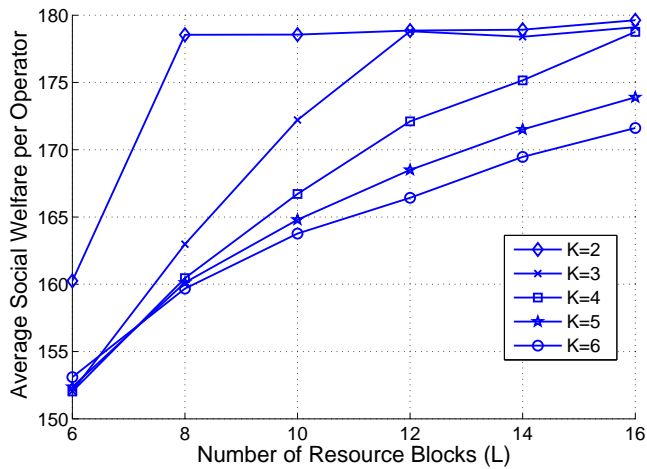}
\caption{Average social welfare per OP for $L=6,8,10,12,14$ and $16$ for different OPs}
\label{AvgwelfareperOP}
\end{figure}

In Fig. \ref{AvgwelfareperOP}, we show the average social welfare per OP (bits/sec/Hz/OP), when there are different numbers of OPs, versus the number of available RBs ($L$). We consider resource demand of each OP to be $c_k = 4$ for all $k \in \Kcal$. The number of OPs that can use the same RB-$l$ is assumed to be $b_l=4$ for all $l\in \Lcal$. For fixed $K$, it can be observed that when we increase the number RBs, the average social welfare per OP increases. Thus, higher number of available RBs will enhance the average social welfare per OP. This is not unexpected as the number of available RBs to be chosen from has increased. However, for fixed $L$, as the number of OPs in the system increases, there is a decrease in average social welfare per OP. This is also not unexpected since increasing the number of OPs, utilizing a fixed number of common RBs, tends to increase the inter-operator interference. 

It is interesting to note that when the number of OPs, $K$, is held fixed, the average social welfare per OP tends to saturate after a certain value of $L$. Increasing the number of RBs does not change the performance anymore. We can easily predict the value of $L$ where this saturation occurs. Since each daughter OP of a parent OP is allocated with orthogonal RBs, inter-operator interference can only occur between the daughter OPs of different parent OP. If the number of RBs is sufficiently large such that each daughter OPs of every parent OPs is allocated RBs orthogonally, then such a saturation occurs, since there is no longer inter-operator interference. For example, since $c_k = 4$ for all $k\in\Kcal$, when there are $K=2$ OPs, the saturation occurs when $L = c_1 + c_2 = 8$. At this point, each OP is allocated with orthogonal RBs. Adding more RBs does not change the orthogonality of the allocation. Similarly, when there are $K=3$, the saturation occurs when $L = c_1 + c_2 + c_3 = 12$; whereas when $K=4$, this happens when $L = 16$.

\subsection{Effect of the Changing the Resource Demand}

\begin{figure}[h]
\centering
\includegraphics[height=5.9in, width=3.8 in, keepaspectratio = true]{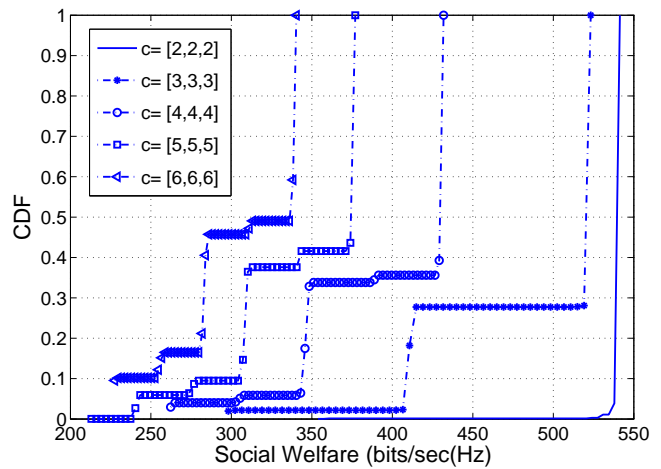}
\caption{Comparison of the cumulative distribution function (CDF) of social welfare for $K=3$ OPs with different sets of $\tbf{c}$ when $L=6$ and $b_l = 4$}
\label{CDFchangecforL6b4}
\end{figure}

Fig. \ref{CDFchangecforL6b4} presents the CDF of the overall social welfare (bits/sec/Hz) when the number of OPs is $K=3$, when the number of RBs is $L=6$, and $b_l =4$ for different sets of $\tbf{c}$. We can observe that when the number of RBs is fixed, with higher resource demand $c_k, \forall k \in K$ of $\tbf{c}$, the CDF curve of the overall social welfare degrades and shifts towards the left. This is because increasing the values of $c_k$ has the effect of increasing the size of the augmented set of OPs. Since each sibling OPs of a parent OP is allocated with orthogonal RBs, more RBs are consumed. However, since the children OPs of other parent OPs need to share the same common resources, this tends to create higher chances of inter-operator interference, when the total number of RB, $L$, is fixed. This leads to an overall decrease in social welfare. For instance, we see that when $\tbf{c} = [2,2,2]$,  the total resource demand $c_1 + c_2 + c_3 = L = 6$. Thus, each OP is allocated with orthogonal RBs, and they are interference free. As the resource demand increases, there are more chances of overlapping assignment of the RBs between the OPs. When $c_k = 6$, each parent OP is assigned with all six available RBs, thus making the system interference limited. In Fig. \ref{CDFchangecforL6b4}, we also observe the dramatic change in the performance of the system when the resource demand is $\tbf{c} = [2,2,2]$ and the system is interference free, to when the system starts to become interference limited when $\tbf{c} = [4,4,4]$.

\begin{figure}[h]
\centering
\includegraphics[height=5.9 in, width=3.8 in, keepaspectratio = true]{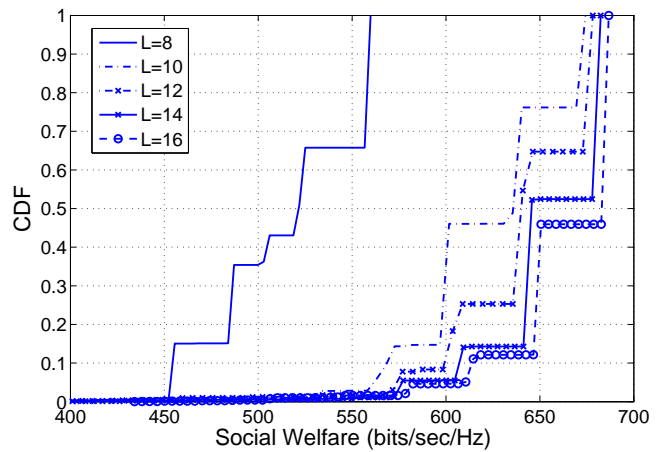}
\caption{Comparison of the cumulative distribution function (CDF) of social welfare for $K=4$ OPs with different $L$ when $b_l = 4$ and $\tbf{c}=[5,9,9,9]$}
\label{CDFchangeLc5999}
\end{figure}

\begin{figure}[h]
\centering
\includegraphics[height=5.9 in, width=3.8 in, keepaspectratio = true]{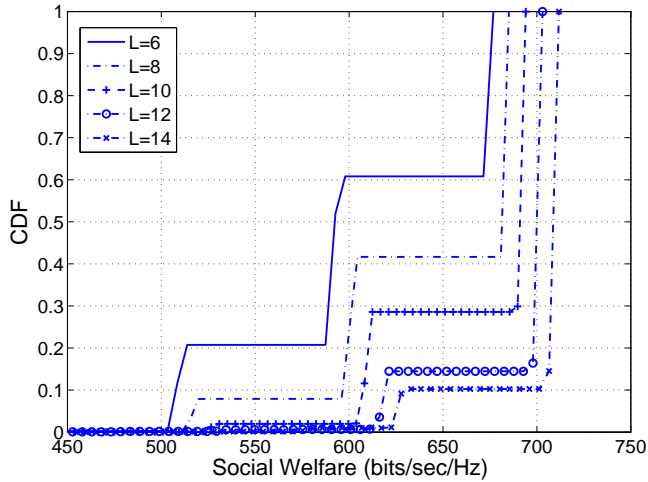}
\caption{Comparison of the cumulative distribution function (CDF) of social welfare for $K=4$ OPs with different $L$ when $b_l = 4$ and $\tbf{c}=[4,4,4,4]$}
\label{CDFchangeLc4444}
\end{figure}

In Fig. \ref{CDFchangeLc5999} and Fig. \ref{CDFchangeLc4444}, we investigate the effect of high and low resource demands by OP as the number of available RBs change. For both the figures, we have fixed the number of OPs to be $K =4$ and the number of OPs allowed to use the same RB-$l$ to be $b_l=4$ for all $l\in \Lcal$. In both the figures, we have plotted the CDF of the overall social welfare (bits/sec/Hz)  when the number of RBs is changed such that $L=8,10,12,14,16$. For Fig. \ref{CDFchangeLc5999}, we have assumed the resource demand by each OP to be high, such that $\tbf{c} =[5,9,9,9]$; whereas for Fig. \ref{CDFchangeLc4444}, we have assumed a much smaller resource demand by each OP, such that $\tbf{c} = [4,4,4,4]$. Both Fig. \ref{CDFchangeLc5999} and Fig. \ref{CDFchangeLc4444} complements the Fig. \ref{CDFchangecforL6b4}, since in both cases, we see the expected increase in social welfare as $L$ is increased, when the resource demand is held fixed. This is because when $L$ increases, there is less chance of inter-operator interference via RB sharing; and the OPs can choose better RBs to maximize their own utilities. Thus, the overall social welfare improves.

In Fig. \ref{CDFchangeLc5999}, since $\tbf{c} = [5,9,9,9]$ represents a high demand, we observe that the CDF of social welfare is enhanced dramatically when the number of RBs is changed from $L=8$ to $L=10$. This is because, when $L=8$,  we have $\sum_{k \in \Kcal} c_{k} = \sum_{l \in \Lcal} b_{l} = 32$. Therefore, when $L=8$, the total supply of resources is equal to the total demand of resources; and the OPs tend to utilize the overall supply of resources. As such, when $L$ is increased, it creates a surplus of resource supply. This means, the number of OPs that need to share the same RB goes down, creating lesser inter-operator interference; and hence, the social welfare tends to enhance significantly.

This is in contrast to Fig. \ref{CDFchangeLc4444}, where the resource demand is assumed to be $\tbf{c} =[4,4,4,4]$, which is much smaller compared to that of Fig. \ref{CDFchangeLc5999}. As such, in Fig. \ref{CDFchangeLc4444}, although the social welfare improves with increasing $L$, we do not see any dramatic improvement in social welfare as witnessed in Fig. \ref{CDFchangeLc5999}.

%
%
\subsection{Effect of the Changing the SBS Intensity}

\begin{figure}[h]
\centering
\includegraphics[height=5.9 in, width=3.8 in, keepaspectratio = true]{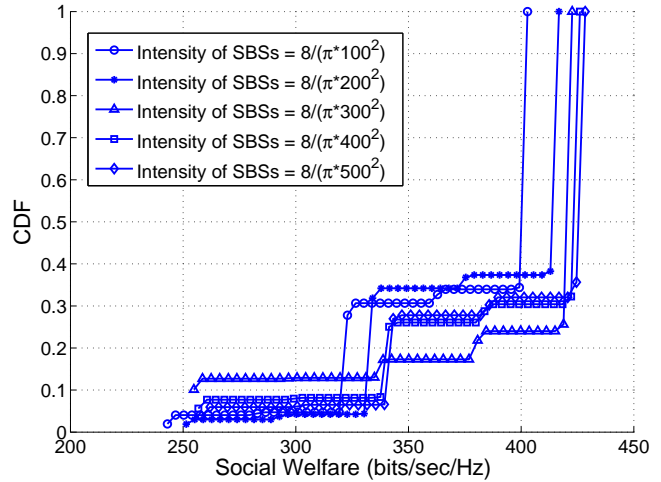}
\caption{Comparison of the cumulative distribution function (CDF) of social welfare for $K=3$ OPs by varying the intensity of SBSs}
\label{CDFchangesradius}
\end{figure}

In Fig. \ref{CDFchangesradius}, we illustrate the effect of changing the SBS intensity on the overall social welfare. We have plotted the CDF of the overall social welfare as the spatial intensity of SBS is changed from $8$ SBSs per $\pi \times 100^{2}$ to $8$ SBSs per $\pi \times 500^{2}$. We consider the number of OPs to be $K=3$ and the number of available RBs to be $L=6$. The number of OPs that can use the same RB-$l$ is assumed to be $b_{l} = 4$, while the resource demand of each OP is assumed to be  $\tbf{c}=[4,4,4]$. The radius of circular area under consideration is varied from $100$ to $500$ meters. Each OP is assumed to serve an 8 SBSs. We can observe that the median of the CDF curves of the social welfare improves as the area increases. This is because when we increase the size of the area, the spatial intensity of the SBS decreases, which leads to less interference.

%
%
%

\section{Conclusion}\label{section:Conclusion}
In this paper, the spectrum assignment for non-orthogonal multi-operator spectrum sharing system, where multiple operators shared a common pool of spectrum among each other, was formulated as a social welfare optimization problem. Using the results from stochastic geometrical analysis, we showed that, under certain condition, the solution to this problem coincided with the solution to a corresponding stable matching game. This result inspired the use of Markov Chain Monte Carlo algorithm to find the stable and socially optimal matchings. The Q-learning method was also proposed to find the optimal random transmit power strategy of the small cell base stations. Numerical simulations were performed to access the performance of the system under various conditions. From the numerical study, one of the conclusions we can draw is that the spectrum allocation has greater effect on the performance of the system than power allocation.

%
%

\bibliographystyle{IEEE}

\end{document}